\documentclass[letter]{aa} 
\usepackage{txfonts,epsfig,graphicx,natbib,url,twoopt}

\newcommand{\omegabold}{\mbox{\boldmath$\omega$}}

\begin{document}

   \title{Returning magnetic flux in sunspot penumbrae}

   \author{B. Ruiz Cobo
          \and
          A. Asensio Ramos
          }

   \institute{Instituto de Astrof\'\i sica de Canarias,
              38205, La Laguna, Tenerife, Spain; \email{brc@iac.es}
            \and
Departamento de Astrof\'{\i}sica, Universidad de La Laguna, E-38205 La Laguna, Tenerife, Spain
             }

  \abstract
   {}
   {We study the presence of reversed polarity magnetic flux in sunspot penumbra.}
   {We applied a new regularized method to deconvolve spectropolarimetric data observed with
   the spectropolarimeter SP onboard Hinode. The new regularization is based on a principal component 
   decomposition of the Stokes profiles. The resulting Stokes profiles were inverted to infer the magnetic field
   vector using SIR.}
   {We find, for the first time, reversed polarity fields at the border of many bright penumbral
filaments in the whole penumbra.}
   {}


   \maketitle
%

\section{Introduction}


MHD simulations of sunspots \citep{rempeletal09a, rempeletal09b, rempel12} have 
successfully reproduced many aspects of sunspot penumbrae as produced by magneto-convection 
in inclined magnetic fields. As a consequence, magnetic field with inverse polarity should be 
observed all along the boundaries of bright penumbral filaments. This reversed flux 
has been observed in the outer penumbra \citep{westend97,westend01a,delToro01}, but even with the 
highest spatial resolution spectropolarimetry, we have been unable to observe them in the inner 
penumbra \citep{langhansetal05}. When inverting data from the Advanced Stokes 
Polarimeter \citep{elmore92} under the MISMA hypothesis, \cite{san05}
found that penumbral filaments could be 
interpreted as bundles of magnetic field lines ranging over all inclination angles. In his PhD thesis, \cite{Franz11} 
shows how the opposite polarity fields in penumbral filaments are responsible for a weak third 
lobe in Stokes $V$ that appears on the red wing of 40\% of the penumbral spectra of the 
two iron lines at 630\,nm observed with the Hinode spectropolarimeter 
\citep[HINODE/SP,][]{litesetal01,kosugietal07}. For recent reviews see \cite{borrero11}, \cite{bellot10}, 
\cite{borrero09}, \cite{sch09}, or \cite{tritschler09}.


\section{Observations}
The active region NOAA 10953 was mapped at an average heliocentric angle of
$\theta$=12.8$^{\circ}$ using HINODE/SP on 2007 Apr 30, between 18:35
and 22:30 UT. This active region has recently been analyzed by
\cite{ichimotoetal09} and \cite{louisetal09, louisetal11}. The region was scanned in $\sim$1000 steps, with a step
width of 0\farcs148 and a slit width of 0\farcs158, recording the full 
Stokes vector of the pair of the neutral iron lines at 630\,nm with a spectral
sampling of 21.46 m\AA. The spatial resolution was $\sim$\,0\farcs32.
The integration time was 4.8 s, resulting in an approximate noise level of
1.2\,$\times$\,10$^{-3}$. To calibrate the spectra, we averaged the intensity
profiles coming from a 5$''\times$5$''$ region of quiet Sun near the disk center. We 
compared this average profile to the FTS spectral atlas
\citep{Kuruczetal84,Brault87} once it was convolved with the spectral point spread function (PSF) of Hinode. We
have found that a veil of 3.4\% must be subtracted from the continuum intensity before normalization. This value is similar to
the stray-light contaminations of 4.7\% found by \cite{danilovic08} and 5\% found by \cite{Socas11}. The 
wavelength was calibrated by
assuming that the average umbral profile is at rest.

\section{Deconvolution}
Deconvolving two-dimensional spectropolarimetric data wavelength by wavelength 
has some drawbacks. First, the number of deconvolutions one has to carry out is large. 
Second, many of these wavelengths contain practically no relevant information 
in Stokes $Q$, $U$ and $V$, apart from the noise. This potentially leads to an enhancement
of the noise level unless it is filtered. For this reason, and in order to
overcome these two problems, we introduced a regularization into the deconvolution
(the details will be given in Asensio Ramos et al. in prep). 
This regularization acts on the spectral dimension of the data by
assuming that the original Stokes profiles (before reaching the telescope) at each pixel can be written as a truncated
linear combination of the principal components $\{\phi_i(\lambda)\}$ obtained from the observations, so that
\begin{equation}
\mathbf{O}(\lambda) = \sum_{i=1}^{N} \omegabold_i \phi_i(\lambda),
\label{eq:pca_decomposition}
\end{equation}
where $N$ is the number of eigenfunctions needed to reproduce the profiles to adequate precision.
Under this assumption, and assuming that the monochromatic PSF is independent of wavelength, the observed
Stokes profiles is given by
\begin{equation}
\mathbf{I}(\lambda) = \sum_{i=1}^{N} \left( \omegabold_i * \mathbf{P} \right) \phi_i(\lambda) + \mathbf{N},
\end{equation}
where $\mathbf{P}$ is the PSF (kindly provided by M. van Noort\footnote{Described in 
\cite{vannoort12} and obtained from the Hinode pupil specified by \cite{suematsu08}.}),
and $\mathbf{N}$ is a Gaussian noise component with zero mean and variance $\sigma^2$. Given the orthonormality of
the principal components, the projection images $\omegabold_i$ fulfill:
\begin{equation}
\langle \mathbf{I}(\lambda), \phi_k(\lambda) \rangle = \omegabold_k * \mathbf{P} + \mathbf{N}.
\end{equation}
where the noise is still Gaussian with zero mean and variance $\sigma^2$.
Consequently, the regularization process we have used implies that we have to deconvolve the projections
of the original measured data onto the basis functions $\phi_k(\lambda)$ and reconstruct
the unperturbed image using Eq. (\ref{eq:pca_decomposition}). One of the most interesting 
properties of this approach is that, since the real signal in each pixel only appears in the first few 
coefficients (a fundamental consequence of the PCA decomposition), the influence of noise is largely minimized. We deconvolve these images
using a Richardson-Lucy algorithm \citep{richardson72,lucy74} controlling the number of iterations to avoid any ringing effect (although we
hardly see any fringing given the absence of noise in the projection images).

\begin{figure}
   \centering
   \includegraphics[width=8cm]{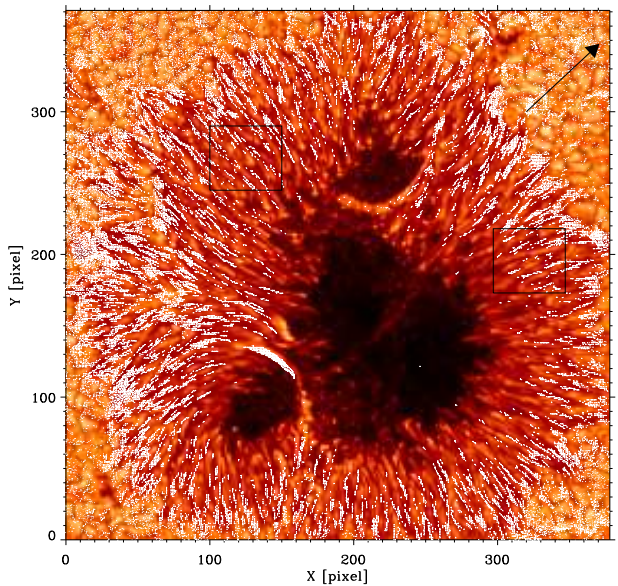}
   \caption{Deconvolved continuum intensity. White arrows correspond to the horizontal
   component of the magnetic field in points with reversed polarity. Black rectangles 
   mark the areas area amplified in Fig. \ref{Fig3}. The black arrow points to the disk center.}
   \label{Fig1}
\end{figure}

\section{Inversion}
The SIR code \citep{ruizcobodeltoro92} was applied on the deconvolved spectra to derive the stratification of 
temperature $T$, magnetic field strength $B$, inclination $\gamma$, azimuth $\phi$, and line of sight 
velocity $V_{\rm los}$ versus continuum optical depth $\tau$.
The values of those parameters were retrieved at a number of optical depth points called nodes. 
We used a maximum of seven nodes in $T$, five in $B, V_{\rm los}$, $\gamma$, and two in $\phi$. 
Five different initializations were used to avoid local minima.
Because we are interested in the penumbra and our sunspot has negative polarity, we solved 
the 180$^{\circ}$ ambiguity of the azimuth by choosing the value in such a way that the magnetic field 
vector \vec{B} points to the umbra. Additionally, we transformed \vec{B} into the local reference frame.
Likewise, we applied the same inversion procedure to the original data, i.e., before
the deconvolution from the spatial PSF. 

\section{Results and discussion}
In Fig. \ref{Fig1} we present the reconstructed continuum intensity map of the region containing the sunspot,
including the horizontal component of the magnetic filed vector in those pixels 
in which the field presents a reversed polarity. We note how the pixels showing reversed polarity are 
distributed over the whole penumbra, with the exception of the most inner part. There is an obvious 
concentration of reversed flux in some patches of the most external part of the outer penumbra. 
The full spot has a total unsigned magnetic flux of $\Phi_z=1.19 \,\rm{x}\, 10^{22}$ Mx, measured 
at $\tau=$0.1, while the flux of the penumbra is  $\Phi_z=0.79 \,\rm{x}\, 10^{22}$ Mx.
The reversed flux in the penumbra accounts for 8.4\% of this penumbral flux, amounting to 17\% of 
the pixels of the penumbra, and 72\% of these pixels present a positive velocity (downflow). 
This figure should be compared with the 56\% of the pixels of the full penumbra harboring downflows. 
The correlation between downflows and reversed fields is probably tighter than this 72\% because 
the velocities are measured in the line of sight reference frame, while \vec{B}
has been translated into the local reference frame.
The preceding figures increase to 30\% (reversed flux over total unsigned flux), with 28\% of the 
pixels harboring reversed polarity and 88\% of these presenting downflows at $\tau=$1.
Nevertheless, \vec{B} at $\tau=$1 has a large uncertainty
(usually error bars around 600 G in $B$ and $15^\circ$ in $\gamma$).

\begin{figure}
   \centering
   \includegraphics[width=8cm]{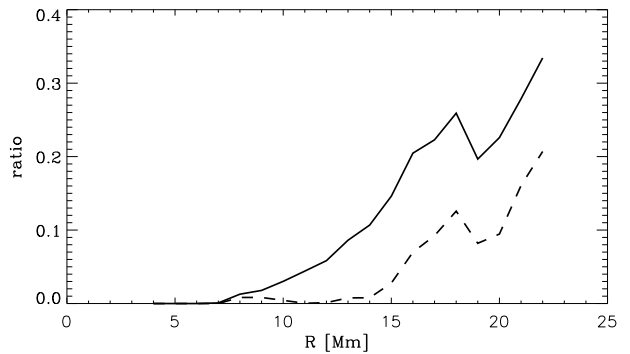}
   \caption{Filling factors of regions with reversed polarity at $\tau=0.1$ after (solid line) and
   before (dashed line) deconvolution.}
   \label{Fig2}
\end{figure}

In Fig. \ref{Fig2} we plot the ratio of pixels harboring inverse polarity with respect to the
total number of pixels at $\tau=0.1$ versus the radius from the center of the umbra, after and before deconvolution.
The ratios shown in this figure are very similar to those found in
numerical simulations \citep[cf. Fig 13 in][]{rempel12}, except
for the increase in the filling factor at radii larger than 19 Mm. We need to
take into account that 
our sunspot is not circular and that the upper left penumbra is certainly wider that
the bottom right one.
However, the similitude with the numerical simulation results is clear, even
considering the difference in spatial resolution. Also the inverse
polarity pixels for the original data 
(before deconvolution) only appear at a radius larger than 14 Mm, while after
deconvolution we already found inverse polarity pixels at 8 Mm.

\begin{figure}
   \centering
   \includegraphics[width=7cm]{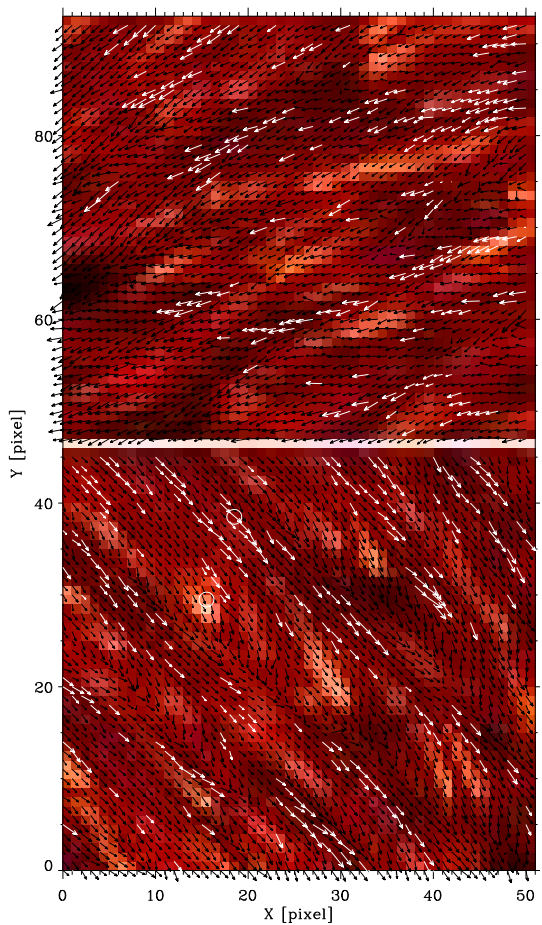}
   \caption{Sections in the center side (upper) and limb side (bottom) penumbra White arrows correspond to the horizontal component of the magnetic 
field in points with reversed polarity. White circles at (15,29) and (18,38) mark the position of the 
spectra shown in Fig. \ref{Fig4}}
   \label{Fig3}
 \end{figure}
 
\begin{figure*}
    \centering
   \includegraphics[width=17cm]{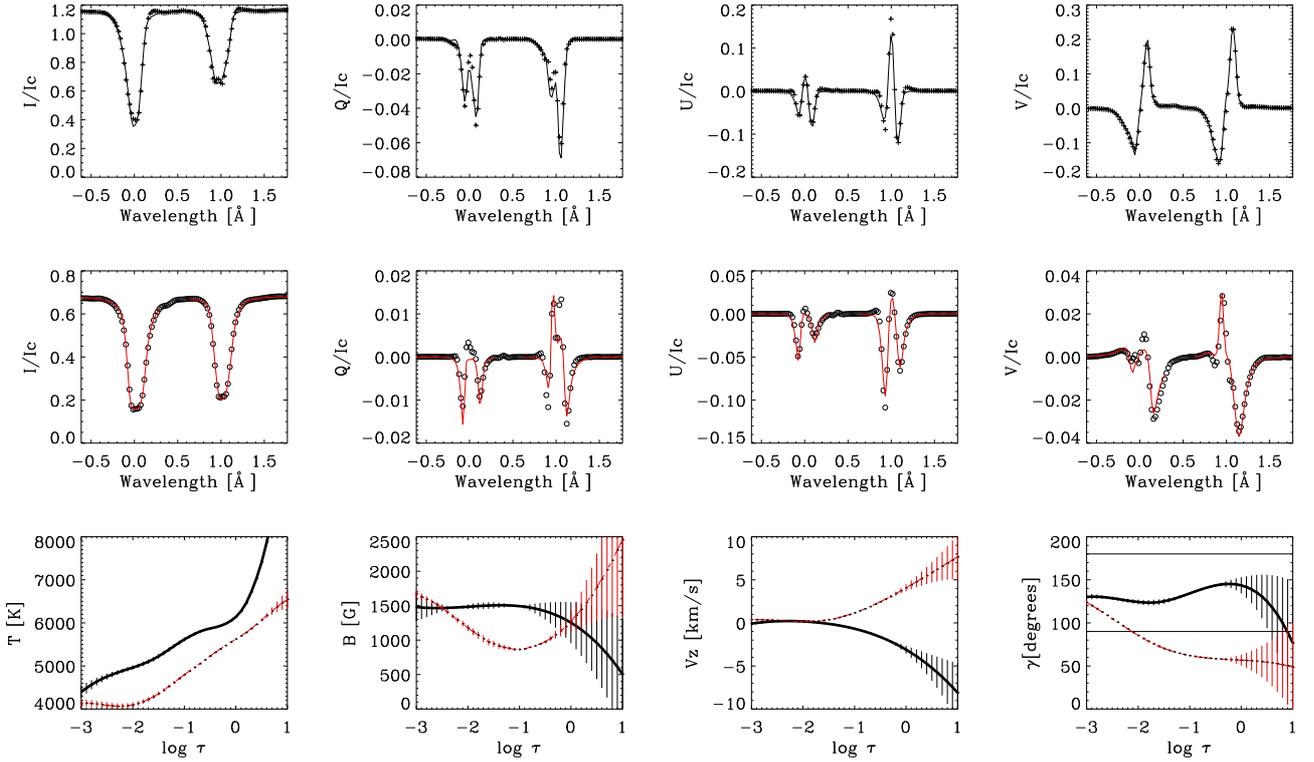}
   \caption{Upper 4 panels: Stokes profiles and best fits obtained by 
the SIR inversion applied to a pixel in a penumbral filament harboring a magnetic field with 
the same polarity of the umbra. Middle 4 panels: the same for a pixel with the reversed polarity.
   Lower 4 panels: corresponding atmospheric models. The thick black line corresponds to the profiles 
shown in the upper row of this figure and the red ones to the middle row.}
   \label{Fig4}
\end{figure*}

In Fig. \ref{Fig3} we show a zoom of two middle penumbra regions (marked with black rectangles 
in Fig. \ref{Fig1}), one on the limb side (bottom panel) and the other on the center side (upper panel).
It is conspicuous that the reversed field pixels are in dark filaments and, in many cases,
they seem to mainly be placed at the boundary of bright filaments.
The circles at (15,29) and (18,38) mark the position of the pixels whose spectra are 
shown in Fig. \ref{Fig4}. 
The first one is an example of a normal polarity pixel, and the second is a reversed one. 
The corresponding stratification of $T, B, V_{\rm z}$, and $\gamma$ in a local reference frame is plotted 
in panels in the bottom row. This pixel harbors an upflow and a negative polarity ($\gamma>90^\circ$). 
The example of a reversed polarity pixel is plotted in the middle row. The corresponding stratification 
through the photosphere is 
plotted in the panels in the bottom row. The Stokes $V$ profiles present a 
clear positive polarity (opposite to that of the umbra). Also we note the small amplitude of these 
Stokes $V$ profiles. The corresponding model is considerably cooler, and it 
presents a downflow and a weaker field with a inclination lower than 90$^\circ$ at all layers
below optical depth 0.01. 
These profiles are contaminated by a small residual of the normal polarity profile. There is 
a small reversed lobe at the blue wing of the Stokes $V$ profiles of both 630.1 nm and 630.2 nm
Fe \textsc{i} lines. This small contamination and the consequent asymmetry of the Stokes profiles 
could be responsible for the strong gradient of both $B$ and $\gamma$,
although the profiles are not perfectly reproduced by the model.

\begin{figure*}
   \centering
   \includegraphics[width=17cm]{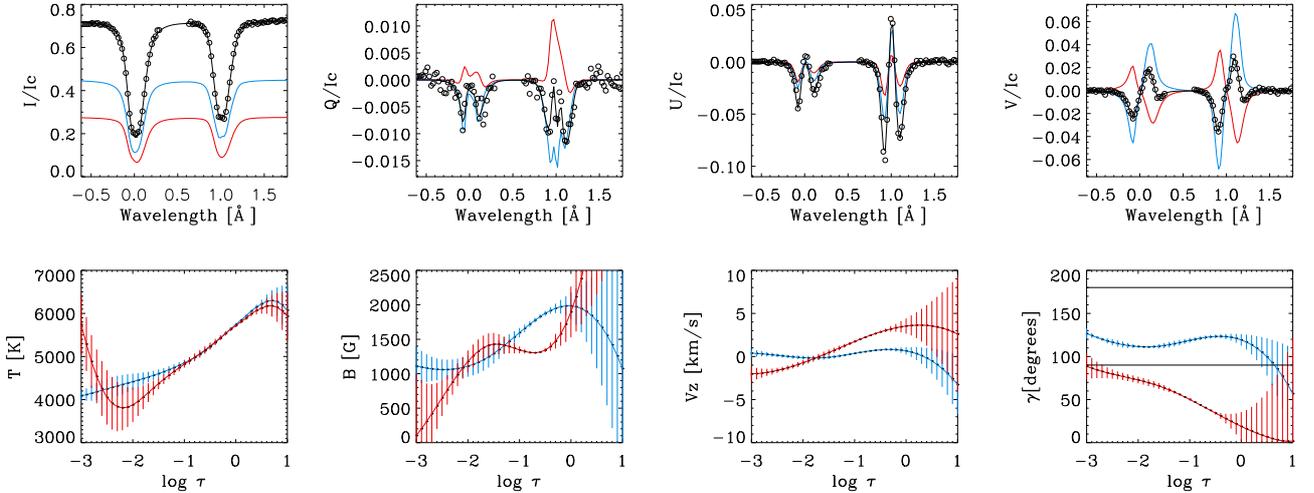}
   \caption{Upper 4 panels: original (before deconvolution) 
Stokes profiles and best fits obtained by a 2-component 
SIR inversion applied to the inverse polarity pixel shown in Fig. \ref{Fig4}. 
The two components are depicted in blue and red, and the combination in black line.
Lower 4 panels: the two atmospheric models.}
   \label{Fig5}
\end{figure*}

It is well known that any overcorrection of stray light in a spectral image generates inverse 
polarity profiles at the border of bright structures. This is caused because a deconvolution 
is equivalent, in those pixels, to the substraction of a fraction of the spectral profile of the 
brighter neighbors. We can assess that our inverse profiles do not come from such an artificial 
effect for two reasons. First, we have undercorrected our data, because
the used PSF is a lower limit of the real one \citep[see, for instance,][]{Joshi11}.   
As a consequence, the continuum contrast has only changed from 6.3\% in the
original data to 11.8\% in the deconvolved data. This value is clearly below
the figure of 13-14\% found in the simulations \citep{danilovic08}. 
Second, we checked that the original (nondeconvolved) Stokes $V$ profiles of the 80\% of those
pixels that harbor an inverse polarity in the deconvolved map present the 
characteristic third lobe in the red wing of Stokes V in the original data 
(37\% of the penumbral pixels harbor such a third lobe in the original data). 
Following \cite{Franz11}, this is a hint 
of the presence of a hidden field with opposite polarity. With these two points we want to demonstrate that 
we are not introducing new information by the deconvolution of the PSF. We are simply disentangling mixed
information. As an illustration, we present in Fig. \ref{Fig5} 
the Stokes profiles before the deconvolution of the same 
inverse polarity profile that we plotted in Fig. \ref{Fig4}. 
Note the presence of the third, inverse polarity, lobe at the red wing of the Stokes $V$ profiles 
of both spectral lines. We inverted these original (nondeconvolved) Stokes 
profiles using SIR, but now considering a two-component atmosphere. In the lower panels one can see the corresponding optical
depth stratification of the different quantities for both models. Interestingly, the model of the 
red component is quite similar to the corresponding model of the deconvolved profile shown in Figure \ref{Fig4}. 
The model of the blue component explains the contamination by stray light,
and consequently it is similar to the model that reproduces a bright pixel. For this reason, the 
blue model is quite similar to the blue model presented in Figure \ref{Fig4}.

\section{Conclusions}
We report here observations of reversed polarity fields at the border of bright penumbral
filaments in the whole penumbra. The spatial distribution of the reversed polarity fields 
is similar to what is predicted by the most recent numerical simulations \citep{rempel12}. 
This has been possible after deconvolution of the original data using a regularization
method based on a principal component decomposition of the profiles.

The Stokes profiles of those pixels that show a reversed polarity in the deconvolved map
display a small opposite polarity third lobe on the red wind in the original data. This is a clear
hint of hidden reversed polarity fields \citep{Franz11}. We found a reversed flux in the 
penumbra at optical depth $\tau=0.1$ of 8.4\% of the total 
unsigned penumbral flux. It corresponds to 17\% of the pixels of the penumbra.
Seventy-two percent of these pixels present positive velocity (downflow). 
All these facts indicate that we have found observational proof 
of convection in sunspot penumbra filaments.

\begin{acknowledgements}
Hinode is a Japanese mission developed and launched by ISAS/JAXA, collaborating with 
NAOJ, NASA and STFC (UK). Scientific 
operation of the Hinode mission is conducted by the Hinode science team organized at 
ISAS/JAXA. Support for the post-launch operation is provided by JAXA and NAOJ (Japan), 
STFC (U.K.), NASA, ESA, and NSC (Norway).
Financial support by the Spanish Ministry of Economy and Competitiveness 
through project AYA2010--18029 (Solar Magnetism and Astrophysical Spectropolarimetry)
is gratefully acknowledged. AAR also acknowledges financial support through the Ram\'on y
Cajal fellowship and the Consolider-Ingenio 2010 CSD2009-00038 project. We thank C. Beck and M. van Noort for fruitful discussions.
\end{acknowledgements}

\end{document}